\documentclass[twocolumn,showpacs,superscriptaddress,prl]{revtex4-1}

\usepackage[dvips]{graphicx}	% Include figure files
\usepackage{dcolumn, bm, color, amsmath}%, cite}
\usepackage{amssymb}

\begin{document}
%\title{High magnetic field studies of the vortex 
%lattice structure in Ca$_{x}$Y$_{1-x}$Ba$_{2}$Cu$_{3}$O$_{7}$}

\title{Probing superconducting order in overdoped Ca$_{x}$Y$_{1-x}$Ba$_{2}$Cu$_{3}$O$_{7}$ by neutron diffraction measurements of the vortex lattice}

\author{A.~S.~Cameron}
\email{alistair.cameron@tu-dresden.de}
\affiliation{School of Physics and Astronomy, University of Birmingham, Edgbaston, Birmingham, B15 2TT, UK.}
\affiliation{Institut f\"ur Festk\"orper- und Materialphysik, Technische Universit\"at Dresden, D-01069 Dresden, Germany.}
\author{E. Campillo}
\email{emma.campillo@sljus.lu.se}
\affiliation{Division of Synchrotron Radiation Research, Lund University, SE-22100 Lund, Sweden}
\author{A. Alshemi}
\affiliation{Division of Synchrotron Radiation Research, Lund University, SE-22100 Lund, Sweden}
\author{M. Bartkowiak}%
\affiliation{%
 Helmholtz-Zentrum Berlin f\"ur Materialien und Energie,Hahn-Meitner-Platz 1, D-14109 Berlin, Germany
}%
\author{L. Shen}
\affiliation{Division of Synchrotron Radiation Research, Lund University, SE-22100 Lund, Sweden}
\author{H. Kawano-Furukawa}
\affiliation{RIKEN Center for Emergent Matter Science (CEMS), Wako, Saitama 351-0198, Japan}
\affiliation{Department of Physics, Advanced Sciences, G.S.H.S. Ochanomizu University, Tokyo 112-8610, Japan}
\author{A.~T.~Holmes}
\affiliation{School of Physics and Astronomy, University of Birmingham, Edgbaston, Birmingham, B15 2TT, UK.}
\affiliation{European Spallation Source ERIC, P.O. Box 176, SE-221 00, Lund, Sweden}
\author{O. Prokhnenko}
\affiliation{%
 Helmholtz-Zentrum Berlin f\"ur Materialien und Energie,Hahn-Meitner-Platz 1, D-14109 Berlin, Germany
}%
\author{A. Gazizulina}
\affiliation{%
 Helmholtz-Zentrum Berlin f\"ur Materialien und Energie,Hahn-Meitner-Platz 1, D-14109 Berlin, Germany
}%
 \author{J. S. White}
 \affiliation{Laboratory for Neutron Scattering and Imaging (LNS), Paul Scherrer Institute (PSI), CH-5232 Villigen, Switzerland}
 \author{R. Cubitt}
 \affiliation{Institut Laue Langevin, 71 Avenue des Martyrs, F-38000 Grenoble cedex 9, France}
  \author{N.-J. Steinke}
 \affiliation{Institut Laue Langevin, 71 Avenue des Martyrs, F-38000 Grenoble cedex 9, France}
\author{C.~D.~Dewhurst}
 \affiliation{Institut Laue Langevin, 71 Avenue des Martyrs, F-38000 Grenoble cedex 9, France}
\author{A.~Erb}
\affiliation{Walther Meissner Institut, BAdW, D-85748 Garching, Germany.}
\author{E.~M.~Forgan}
\affiliation{School of Physics and Astronomy, University of Birmingham, Edgbaston, Birmingham, B15 2TT, UK.}
\author{E.~Blackburn}
\affiliation{School of Physics and Astronomy, University of Birmingham, Edgbaston, Birmingham, B15 2TT, UK.}
\affiliation{Division of Synchrotron Radiation Research, Lund University, SE-22100 Lund, Sweden}

\begin{abstract}
We present small angle neutron scattering studies of the magnetic vortex lattice (VL) in Ca$_{0.04}$Y$_{0.96}$Ba$_{2}$Cu$_{3}$O$_{7}$ up to a field of 16.7 T, and Ca$_{0.15}$Y$_{0.85}$Ba$_{2}$Cu$_{3}$O$_{7}$ up to 25 T. We find that the series of vortex lattice structure transitions have shifted down in field relative to those reported for the undoped compound.  We attribute this mainly to the weakening of the 1-D superconductivity in the Cu-O chains by the disorder introduced by doping. The hole doping by calcium is also expected to alter the Fermi velocity and it reduces the upper critical field of the system. The high-field structure of the vortex lattice is similar to recent measurements on the parent compound in fields of 25~T, which indicates that the fundamental \textit{d}-wave nature of the superconducting gap is unchanged by calcium doping. This is corroborated by the temperature dependence of the VL form factor which also shows the same \textit{d}-wave behaviour as observed in other cuprates. We also find evidence of Pauli paramagnetic effects in the field dependence of the VL form factor.
\end{abstract}

%\pacs{
%74.25.Wx, % Vortex lattices, flux pinning and creep
%74.72.Gh, % Y-based s/c 
%28.20.Cz % Neutron diffraction and SANS
%}

% PACS, the Physics and Astronomy Classification Scheme.

\date{\today}

\maketitle

\section*{Introduction}

\noindent We report on small angle neutron scattering (SANS) studies of the magnetic vortex lattice (VL) in  Ca$_{x}$Y$_{1-x}$Ba$_{2}$Cu$_{3}$O$_{7}$ (Ca-YBCO), with $x=0.04~\&~0.15$. The parent compound of Ca-YBCO is YBa$_{2}$Cu$_{3}$O$_{7}$ (YBCO$_{7}$), the fully oxygen doped member of the YBa$_{2}$Cu$_{3}$O$_{7-\delta}$ (YBCO) family. YBCO was the first high $T_{\rm c}$ superconductor where the mixed state was studied by SANS \cite{For90}, and since then, VL studies using SANS have continued to provide a wealth of information about its superconducting state ~\cite{For90,Yet93a,Yet93b,Kei93,Kei94,For95,Aeg98,Joh99,Bro04,Sim04,Whi08,Whi09,Whi11,Cam14}. Typically, hole doping in YBCO is performed by varying oxygen content on the Cu-O chains. The fully oxygenated compound  with $\delta = 0$ is very slightly overdoped, but has low pinning due to the absence of oxygen vacancies. To increase the hole concentration further, the system can be doped with calcium, which has one less electron in its outer orbital structure than yttrium. Hole doping in this manner, unlike oxygen doping, does not modify the structure of the copper oxide chains, allowing us to  maintain fully-occupied chains, while altering hole doping in the planes. The hole contribution of the Cu-O chains is not trivially related to the oxygen content, being strongly affected by disorder in the chains~\cite{Jor90}. Using a phenomenological relation between the critical temperature in YBCO and the hole doping $p$~\cite{Tal95}, we can estimate that the sample with $x = 0.04$ ($T_{\rm c} = 79$~K~\cite{Ber96}) has $p = 0.20$, while the $x = 0.15$ sample ($T_{\rm c} = 57$~K~\cite{Ber96}) has $p = 0.23$. It is also suspected that the chains themselves become superconducting through the proximity effect, potentially adding an \textit{s}-wave admixture to the order parameter of the system~\cite{Kir06}. Furthermore, the contribution to the superconducting order parameter from the Cu-O chains behaves differently in YBCO to those in the double-chained YBa$_{2}$Cu$_{4}$O$_{8}$ system \cite{Cam14, Whi09, Whi11, Whi14}. 
%In addition to the complications of chain superconductivity, the doping level itself is expected to influence the order parameter in the cuprates \cite{Yeh01, Deu99}. The VL is sensitive to the structure of the superconducting order parameter, providing a bulk probe with which to investigate this question.
In addition to the complications of chain superconductivity, the doping level itself is expected to influence the order parameter in the cuprates \cite{Yeh01, Deu99}.  Since the VL is sensitive to the structure of the superconducting order parameter, it is an excellent probe to investigate these matters.

\section*{Experimental Details}

\begin{figure*}[th!]
\begin{center}
	\includegraphics[width=1\linewidth]{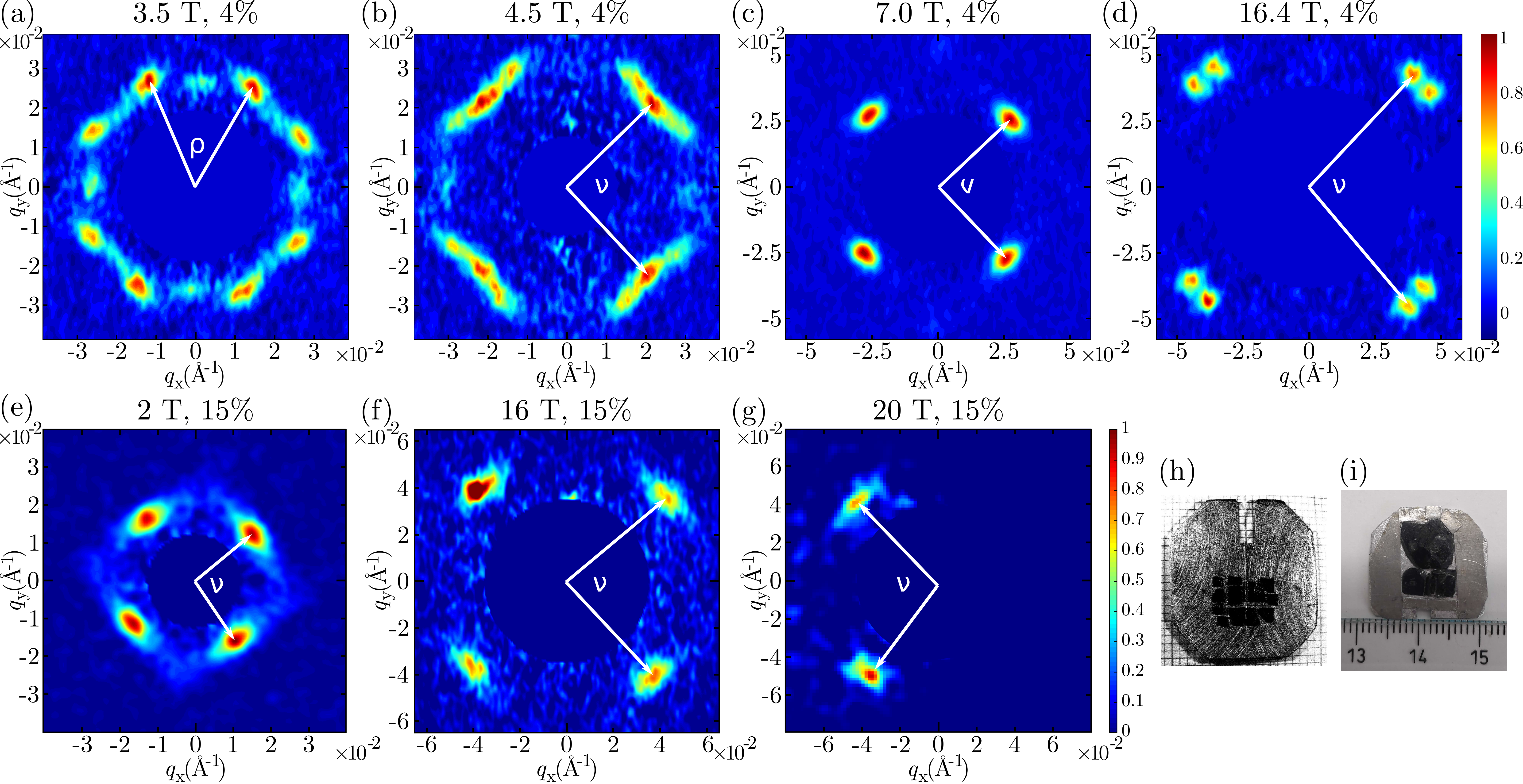}
\end{center}
	\caption{(Color online) Diffraction patterns from both the 4\% calcium doped sample (a -- d) and the 15\% calcium doped sample (e -- g), in applied fields of (a) 3~T, (b) 4.5~T, (c) 7~T, (d) 16.4~T, (e) 2~T, (f) 16~T and (g) 20~T. In panels (e) and (f) the sample was slightly rotated around the $\bm{c}$ axis as compared to the other panels, so that the spots are not exactly centered on $45^{\circ}$ to the horizontal axis.  In panel (g) measurements were done only for the Bragg condition on the -$q_x$ side. The opening angle of the VL is defined as $\rho$ for the hexagonal lattice and $\nu$ for the rhombic lattice, and diffraction patterns are plotted on a normalised intensity scale. (h) Photograph of the 4\% doped sample. (i) Photograph of the 15\% doped sample.}
		\label{Graph1}
\end{figure*}

\noindent
The samples were (i) a mosaic of 12 lightly-twinned single crystals of Ca-YBCO with 4\% calcium doping, which had been oxygenated to O$_7$ under high pressure, so that the Cu-O chains were complete; (ii) a mosaic of 3 more-heavily-twinned single crystals of Ca-YBCO with 15\% calcium doping, which had similarly been oxygenated to O$_7$. The crystals were mounted on plates of high purity aluminum, co-aligned so that the \textbf{a} and \textbf{b} axes were horizontal or vertical and the \textbf{c} axis of the crystals were perpendicular to the sample plate. Measurements were taken with the \textbf{c}-axis of the crystals at 10$^{\circ}$ to the applied field, to reduce the pinning of the flux lines to twin planes~\cite{Bro04}. This value of angle was chosen to noticeably reduce vortex pinning, but so that the effect of the tilted field on the superconducting properties experienced by the VL was small. (The superconducting properties are expected to vary as the cosine of the angle between the \textbf{c}-axis and applied magnetic field, which in this case is $> 0.98$ compared with 1.0 for zero tilt~\cite{Cam88}.) These measurements were performed on the D33 instrument~\cite{Dew08} at the Institut Laue-Langevin (ILL) in Grenoble, France \cite{ILL2013,CampilloILL2021}, using the Birmingham 17~T cryomagnet~\cite{Hol12}, the SANS-I instrument at the Paul Scherrer Institut (PSI) Villigen, Switzerland, and at the HFM/EXED beamline at HZB, Berlin \citep{Pro15,Pro17,Sme16}. 

The data from ILL and PSI were collected in monochromatic mode.  The VL was prepared using the oscillation-field-cool method, whereby a small oscillation, on the order of 1\%, was made in the magnitude of the applied field as the sample was cooling to base temperature.   The resulting data were analysed using the GRASP analysis package \cite{GRASP} and the diffraction patterns, such as those in Fig.~\ref{Graph1}(a)-(f), were treated with a Bayesian method for analysis~\cite{Hol14}. After cooling, the field was held fixed during the measurements, and for the scans as a function of temperature, data were taken on warming. Measurements were performed by rotating the sample and applied field together through the Bragg conditions for the vortex lattice diffraction spots, with background measurements taken above $T_{\rm c}$ and subtracted from the low temperature data to leave only the signal from the vortex lattice.

For measurements performed on the HFM/EXED instrument we relied on the small ripple from the hybrid magnet to provide the oscillation during cooling in field, rather than direct control. This instrument operated using the time-of-flight (TOF) technique \cite{Cam22,Cam22TOF} which requires one or two fixed magnet rotation angles, for foreground and similarly for background. These data were analysed using the Mantid \cite{Arn14} software package.   

\section*{Results and Discussion}

\subsection*{Field dependence of the vortex lattice structure}

\begin{figure}
	\includegraphics[width=1\linewidth]{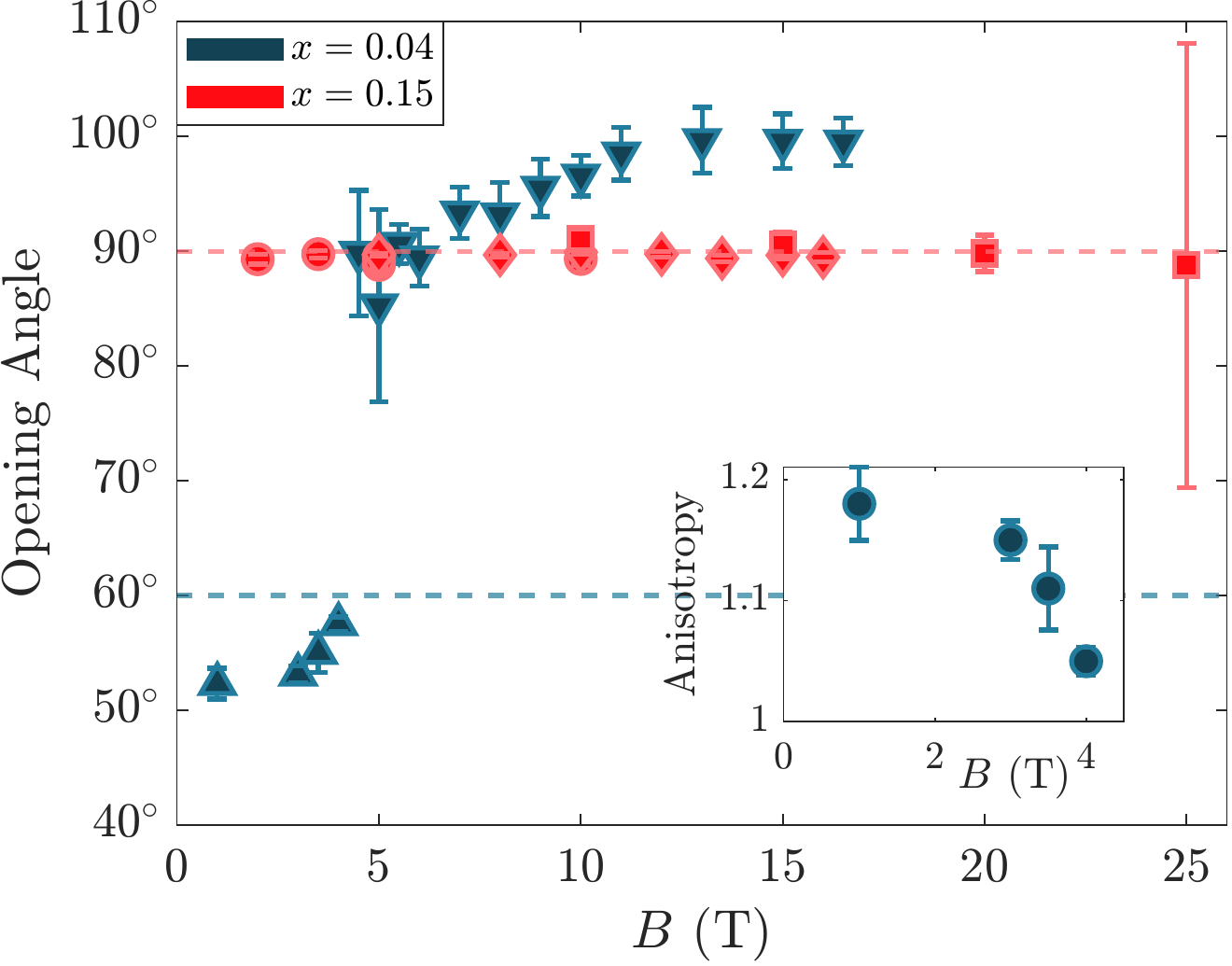}
	\caption{(Color online) Opening angle of the VL as a function of applied magnetic field. Dashed lines indicate the opening angle for a hexagonal and square lattice, with the respective angles for the measured VL, $\rho$ and $\nu$, defined in Fig.~\ref{Graph1}. $x$ refers to the Ca-doping level in Ca$_{x}$Y$_{1-x}$Ba$_{2}$Cu$_{3}$O$_{7}$. Dark blue ($\triangle$) and ($\triangledown$) symbols correspond to the data obtained at ILL in 2016 for Ca$_{0.04}$Y$_{0.96}$Ba$_{2}$Cu$_{3}$O$_{7}$. The dark blue ($\circ$) symbols (inset) correspond to the anisotropy of the VL (defined in the main text). Red points correspond to data from Ca$_{0.15}$Y$_{0.85}$Ba$_{2}$Cu$_{3}$O$_{7}$; ($\circ$) data were obtained at PSI in 2021, the ($\diamond$) symbols correspond to data measured at ILL in 2021 and the red ($\square$) symbols are from HFM/EXED in 2019.}
		\label{Graph2}
\end{figure}

\noindent
Typical VL diffraction patterns from both 4\% and 15\% doped samples are presented in Fig.~\ref{Graph1}, illustrating the various VL structures observed as a function of field. From the 4\% doped sample, in panel (a) at 3.5 T we see twelve spots, corresponding to two distorted hexagonal VLs which form in the two domains of the twinned crystals. By comparison of these results with the data from YBCO$_7$ at 5 T, presented in Ref.~\cite{Whi11}, we can identify the present VL structure with that seen between 2.5 \& 6.5~T in the parent compound. That showed weak diffraction spots along the {\bf b*}-axis, and four stronger off-axis spots. Hence the VL spots bound by the angle $\rho$ in Fig.~\ref{Graph1}(a) correspond to the crystal domain with {\bf b*} horizontal, along which direction are the two weaker spots from this VL. 
%The distortion of this VL away from regular hexagonal is consistent with a larger superfluid density along the {\bf b}-axis, arising from superconductivity of the carriers in the chains~\cite{Whi11}. 
For a  regular hexagonal VL, the diffraction spots lie on a circle, while for a distorted VL they lie on an ellipse, with the amount of distortion characterized by the axial ratio. As in YBCO$_7$ the distortion is consistent with a larger superfluid density along the {\bf b}-axis, giving anisotropy in the London penetration depth, arising from superconductivity of the carriers in the chains~\cite{Whi11} .
We find that the anisotropy is more strongly suppressed by field than in YBCO$_7$~\cite{Whi11} and is nearly absent by 4 T (see inset in Fig.~\ref{Graph2}). At 4.5~T (Fig.~\ref{Graph1}(b)), the VL appears somewhat disordered, and mixed-phase. We believe this indicates that the VL structure undergoes a first order transition from the low field hexagonal lattice (Fig.~\ref{Graph1}(a)) to a high field rhombic lattice at 4.5~T (Fig.~\ref{Graph1}(b-d)).  Again, by comparison with the results of Ref.~\cite{Whi11}, we can identify this with the rhombic structure which appears at 6.5~T and above in YBCO$_7$. In our twinned sample, we would expect two domains to be present. If so, the spots are too close to be clearly resolved when the phase first appears. Alternatively, in this field-region, the VL structure may be pinned to the twin planes, which would give spots at $45^{\circ}$ to the crystal axes. At first glance, it appears that between 4.5~T and 8~T there is a square phase, shown in Fig.~\ref{Graph1}(c), which then undergoes a transition to the rhombic phase, which is clearly present in panel~(d). However, upon closer inspection it is seen that the VL diffraction spots in the ``square'' phase elongate tangentially with increasing field before separating. This indicates that the higher field region is a single rhombic phase, with a lattice close to square at fields just above the transition, and continuous evolution of its structure with field causing the diffraction spots from the two different domains to separate at higher fields. In the 15\% doped sample, in panels (e--g), it appears that only the rhombic high-field phase is present. 

The VL structures as expressed by the opening angles $\rho$ and $\nu$, defined in Fig.~\ref{Graph1}, are shown in Fig.~\ref{Graph2}. For the 4\% sample, the low field hexagonal phase is slightly distorted, with an opening angle $\rho$ of less than $60^{\circ}$. The first order transition to the rhombic phase is seen at 4.5~T, after which the lattice structure evolves smoothly with increasing field until around 13~T. From here we see that the structure remains constant to the highest applied field of 16.7 T, with an opening angle $\nu$ of around $100^{\circ}$. By comparison of the field-dependence of $\nu$ with that seen in the undoped compound, we expect that the rhombic spots labeled by $\nu$ in Fig.~\ref{Graph1}(d) arise from the VL in the crystal domains that have {\bf b*} horizontal, and the spots closer to the horizontal arise from the VL in the crystal domains that have {\bf b*} vertical. A low-field hexagonal to a high-field rhombic VL transition is characteristic of the cuprates, and is taken to be an indication of the predominantly \textit{d}-wave order parameter. In comparison with the parent compound, however, we only observe two of the three structure phases found in YBCO$_{7}$ \cite{Whi09, Whi11}, which exhibits two hexagonal (low-field and mid-field) and one rhombic (high-field) VL structure phases. The single hexagonal phase we have observed in Ca-YBCO is analogous to the mid-field hexagonal phase of YBCO$_{7}$.  However, we cannot rule out the presence of the low-field phase in Ca-YBCO, because the VL diffraction pattern is too disordered below an applied field of 2~T to determine the VL coordination and orientation precisely.  Returning to the mid-field hexagonal to rhombic transition in 4\% Ca-YBCO: this takes place at a lower field than the same transition in YBCO$_7$, and a field-independent value of $\nu$ above $\sim 13$~T seen here is not observed within the same field range in the parent compound. Rather, the rhombic phase in YBCO$_{7}$ appears to be evolving towards a field-independent $\nu$ at around 23-25~T \cite{Cam22}. These observations indicate that the VL in 4\% Ca-YBCO possesses a similar phase diagram to its parent compound, but shifted down in field by approximately a factor of 2. For the 15\% sample, the opening angle \emph{appears} in Fig.~\ref{Graph2} to remain at $90^{\circ}$, but the spots in Fig.~\ref{Graph1}(e)-(g) are elongated tangentially, like those in Fig.~\ref{Graph1}(c). We take this to indicate that in this sample too, the true value of $\nu$ would be $> 90^{\circ}$, but this is obscured by the stronger effects of twin-plane pinning in the 15\% sample.

The VL structure transitions observed here, and in other systems, can be attributed to anisotropies in both the electronic structure and the superconducting gap. In general, VL structure theories attempt to model this  either by considering an anisotropic Fermi velocity in the presence of an isotropic superconducting gap, or vice versa \cite{Kog97a, Kog97b, Kog96, Fra97, Aff97, Ami98, Ich99, Suz10, Ami00, Nak02}. We find several problems when drawing comparisons between these models and our results. Firstly, these theories tend to discuss tetragonal systems, whereas ours is orthorhombic. Consequently, they predict a high-field  square VL structure, rather than the rhombic lattice we find here. Secondly, as was noted in work on YBCO$_{7}$~\cite{Whi11, Whi09}, they predict a single 45$^{\circ}$ rotational transition between two distorted hexagonal phases rather than the 90$^{\circ}$ transitions observed in the YBCO compounds discussed here. However, beyond this we are able to draw qualitative comparisons between the theory and our results. In both the $\beta$ model of Suzuki~\textit{et al}.~\cite{Suz10} and the model of Affleck~\textit{et al}.~\cite{Aff97}, which respectively consider  anisotropy in Fermi velocity and the superconducting gap, the transition to the high-field structure moves to lower fields as the anisotropy in either the Fermi velocity or the gap is increased. This suggests these anisotropies are more pronounced in Ca-YBCO. There are also first-principles calculations using Eilenberger theory, which have predicted that at high field the vortex nearest-neighbor directions align along the nodes of the order parameter~\cite{Ich99}. Our observation that - in the high-field limit of our experiments - the VL nearest-neighbor directions are essentially the same in 4\% Ca-YBCO and YBCO$_7$ indicates that calcium doping leaves the superconducting gap node directions little changed in this region. However, increased hole doping does cause the upper critical field to fall  \cite{Gri14}, so it seems likely that the addition of calcium is reducing the field-scale by this means. We note that there is contradictory evidence in the literature regarding the influence of calcium doping on the hole content of the CuO$_{2}$ planes. Raman spectroscopy measurements suggest that Ca atoms form independent nanophases in the material, which would mean that additional carriers from the calcium do not contribute to doping \citep{Lia05a}, whereas scanning tunnelling spectroscopy measurements have indicated that calcium doping does contribute additional holes to the system \cite{Yeh01}. Our measurements, showing clear differences in the field-scale of VL transitions in YBCO$_{7}$ and Ca-YBCO, support the latter scenario. However, characteristic fields can be identified for both the YBCO$_{7}$ and Ca-YBCO, in particular the field at which the opening angle stops changing, and the point at which the opening angle goes through 90$^{\circ}$.  The fields for the 4\% doping are approximately a factor of two smaller than in YBCO$_7$. This is larger than expected given the different $T_c$ values and the effect of hole-doping on the critical field \cite{Gri14}.  Indeed, Grissonnanche \textit{et al.} \cite{Gri14} find that a 5\% doped sample has an upper critical field that is two-thirds that of YBCO$_7$.

\begin{figure}[b!]
    	\includegraphics[width=\columnwidth]{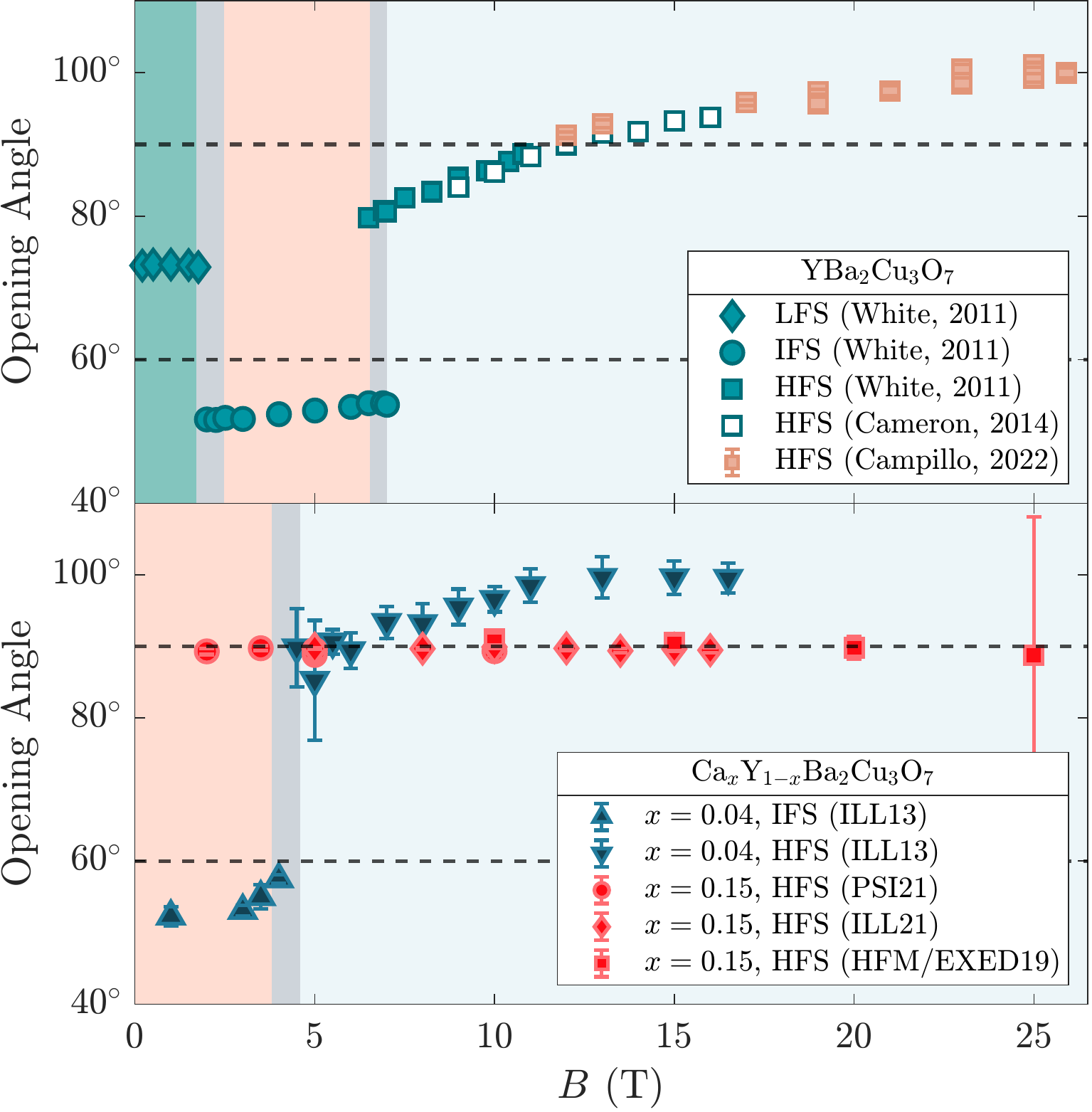}
 	\caption{(Color online) Opening angle of the VL as a function of applied magnetic field for YBa$_2$Cu$_3$O$_7$ and Ca$_{x}$Y$_{1 - x}$Ba$_{2}$Cu$_{3}$O$_{7}$. Dashed lines indicate the opening angle for a hexagonal and square lattice. Meanwhile YBa$_2$Cu$_3$O$_7$ passes through a low field structure (LFS), an intermediate field structure (IFS) and stabilizes in a high field structure (HFS), Ca$_{0.04}$Y$_{0.96}$Ba$_{2}$Cu$_{3}$O$_{7}$ goes through the IFS and also stabilizes in the HFS and Ca$_{0.15}$Y$_{0.85}$Ba$_{2}$Cu$_{3}$O$_{7}$ remains at the HFS for the entire field range.}
		\label{Graph3}
\end{figure}

\subsection*{Field dependence of VL form factor}

\begin{figure}
	\includegraphics[width=1\linewidth]{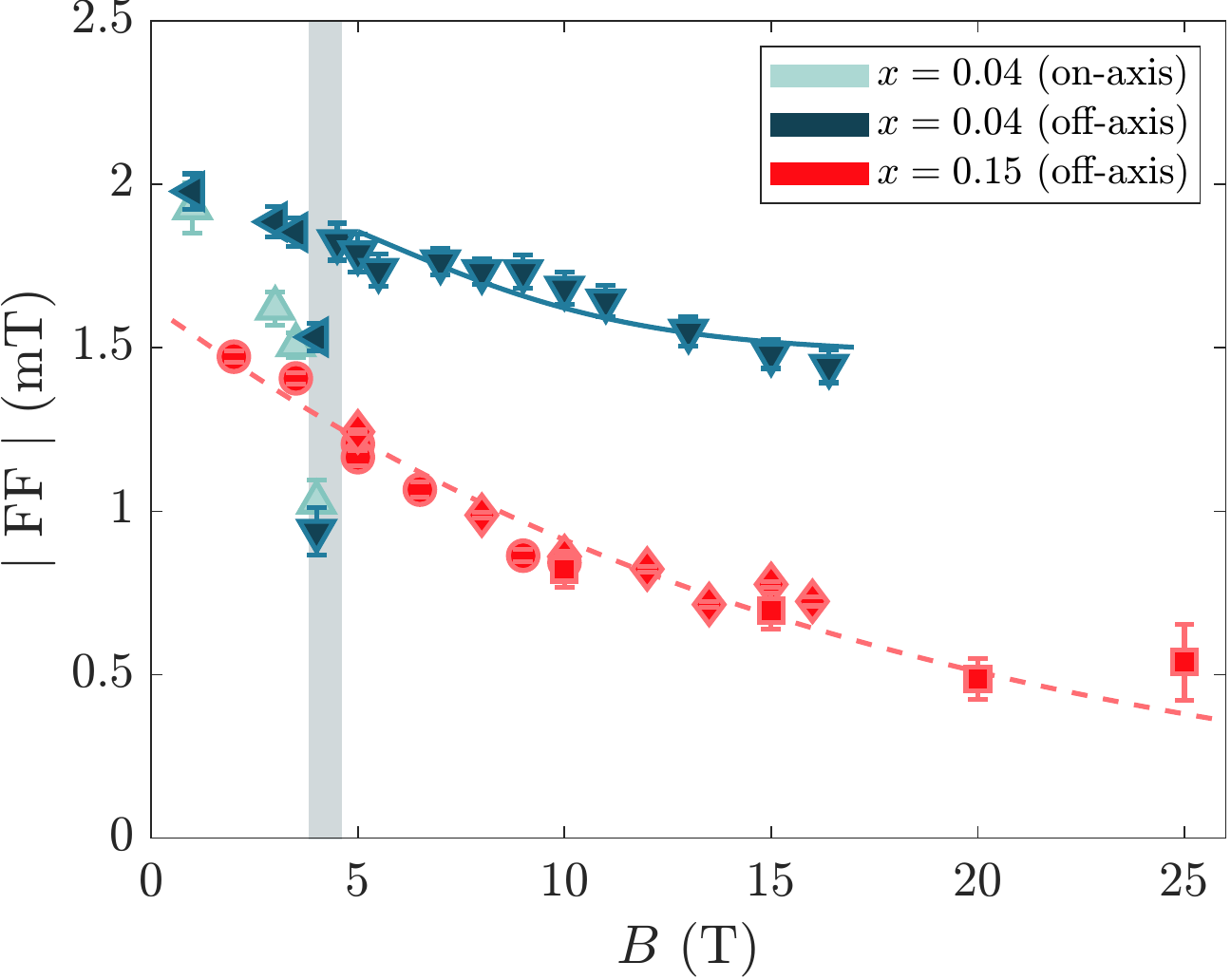}
	\caption{(Color online) Vortex lattice form factor as a function of applied magnetic field. $x$ refers to the Ca-doping level in Ca$_{x}$Y$_{1-x}$Ba$_{2}$Cu$_{3}$O$_{7}$. The shaded grey area defines the phase transition between low- and high-field phases for the 4\% sample, which is observed in Fig.~\ref{Graph1}(b), where both phases are present. The solid line is a fit of the 4\% high field data to the model of equation~\ref{LondonFF} and the lower dashed line is a fit to the 15\% data using the extended London model. Dark blue ($\triangle$) and ($\triangledown$) symbols correspond to data obtained at ILL in 2013 for Ca$_{0.04}$Y$_{0.96}$Ba$_{2}$Cu$_{3}$O$_{7}$. 	Red points correspond to data from Ca$_{0.15}$Y$_{0.85}$Ba$_{2}$Cu$_{3}$O$_{7}$; the ($\circ$) data were obtained at ILL in 2021, the ($\diamond$) symbols correspond to data measured at ILL in 2021 and the ($\square$) symbols are from data obtained at HFM/EXED in 2019. 
	%The grey symbols correspond to previous SANS measurements performed on YBa$_{2}$Cu$_{3}$O$_{7}$ in the high field region \cite{Whi11,Cam14,Cam22}.
	}
		\label{Graph4}
\end{figure}

\noindent
The spatial variation of magnetic field within the vortex lattice, expressed by the VL form factor, is shown in Fig.~\ref{Graph4} for the diffraction spots seen in  Fig.~\ref{Graph1}. The form factor is a Fourier component of the field variation at the scattering vector $\textbf{q}$ , and is related to the integrated intensity, $I(\textbf{q})$, as measured by SANS, through the relation \cite{Chr77}:
\begin{equation}
I_{\textbf{q}} = 2\pi V\phi \big(\frac{\gamma}{4} \big)^{2} \frac{\lambda_{n}^{2}}{\Phi_{0}^{2}q} |F(\textbf{q})|^{2}.
\label{FF}
\end{equation}
Here $V$ is the sample volume, $\phi$ is the flux of incident neutrons, $\gamma$ is the magnetic moment of the neutron in nuclear magnetons (1.91), $\lambda_{n}$ is the wavelength of the incident neutrons and $\Phi_{0}$ is the flux quantum, $h/2e$. The spatial variation of field within the mixed state is determined both by the London penetration depth $\lambda$ and the coherence length of the Cooper pairs $\xi$. As remarked earlier, the field direction is sufficiently close to the {\bf c}-axis that the response is dominated by the carrier motion in the basal plane. Hence we can use an expression for an orthorhombic superconductor with the field applied parallel to the \textbf{c}-axis of the material, with the form factor given by the anisotropic extended London model (ALM)~\cite{Cle75, Hao91, Yao97}:
\begin{equation}
F(\textbf{q}) = \frac{\langle B\rangle \exp(-c(q_{x}^{2}\xi_{b}^{2} + q_{y}^{2}\xi_{a}^{2}))}{q_{x}^{2}\lambda_{a}^{2} + q_{y}^{2}\lambda_{b}^{2}}
\label{LondonFF}
\end{equation}
where $\langle B \rangle$ is the average internal induction, $\xi_{i}$ is the coherence length along axis $i$, $\lambda_{i}$ is the penetration depth arising from supercurrents flowing in direction $i$, and $q_{x}$, $q_{y}$ are in-plane Cartesian components of the scattering vector, with $q_{x}$ parallel to \textbf{b}$^{\ast}$. The parameter \textit{c} accounts for the finite size of the vortex cores, and a suitable value for $c$ in our field and temperature range is 0.44 \cite{Cam22}. For fields close to $B_{\textrm c1}$, the denominator needs an additional $+1$, but for the fields employed here, this addition is negligible.

We find, however, no values of $\lambda$ and $\xi$ for which equation~\ref{LondonFF} is able to fit the full range of 4\% data in Fig.~\ref{Graph4}. The fit presented in this figure is for the high-field phase only.  We use a modified anisotropic London model as in Ref.~\cite{Cam22}, with the basal-plane penetration depths linked as follows:
\begin{equation}\label{fourth}
\lambda_{b}^{2}(B)  =  \lambda_{a}^{2}\lbrace 1 + 0.4 \cdot \tanh\left[ (B-5~\mathrm{T})/7~\mathrm{T} \right] \rbrace.
\end{equation}
This equation differs from that used in Ref.~\cite{Cam22}; the cross-over field of 5 T is half that used for YBCO$_7$. 

This fit returns a value for $\lambda_{\rm a} = 168(3)$~nm.  The value for the London penetration depth seems quite reasonable, being slightly higher than for the parent compound YBCO$_{7}$. The fitted coherence length would be the average of that in the $a$ and $b$ directions (because the Bragg reflections are close to $\lbrace110\rbrace$ directions), but it is unphysically small (0.14 \AA).  On the other hand, assuming that the field-scale has been suppressed still further, the 15\% Ca-doped form factor was fitted with the a constant penetration depth to the extended London model, giving reasonable $ab$-average values of $\lambda = 179(3)$~nm and $\xi =$ 2.64(8)~nm ($B_{\rm c2} = 47(3)$~T).

This deviation of the form factor from the extended London model was also observed within previous SANS studies on the parent compound~\cite{Whi11,Cam14,Cam22}, although this began at the much higher field of $\sim 12$~T, and continued through to the highest measured field of 25~T. In the original study where the deviation of the form factor from the model was reported, it was suspected that disorder in the vortex lattice was contributing to a static Debye-Waller effect, which reduced scattering from the VL at lower fields~\cite{Cam14}. With increasing field, the corresponding increase in the inter-vortex interaction was proposed to overcome the pinning of the flux lines to defects in the crystal lattice, reducing the static Debye-Waller effect and leading to the apparent increase in the VL form factor which rendered the London model unable to fit the data. Corroborating evidence for this was seen in the temperature dependence of the VL form factor, where indications of an irreversibility temperature suggesting the crossing of a glass-solid transition were seen. However, in the temperature dependent data presented in Fig.~\ref{Graph6}, which will be discussed in detail later, we see no indication of such an irreversibility temperature. 

We therefore turn to the possibility of the Pauli paramagnetic effect, whereby the Fermi surface splitting of spin up and spin down electrons by the Zeeman effect leads to the paramagnetic breaking of Cooper pairs. Pauli paramagnetic pair breaking is one of two possible mechanisms by which superconductivity is destroyed at $H_{\rm c2}$, so at such low fields it is not expected to be relevant in the bulk. However, within the vortex core region the Cooper pairs are much less strongly bound, and so this can lead to the formation of a paramagnetic moment and the corresponding alteration of the vortex core structure~\cite{Ich07,Campillo2021}. This would increase the field contrast between the cores and the bulk which is observed as a corresponding increase in the VL form factor. This has also been observed in heavy-fermion superconductors~\cite{Bia08, Whi10,Campillo2021}, a borocarbide~\cite{DeB07} and an iron-based superconductor~\cite{Kuh16}. We have speculated that in YBa$_2$Cu$_3$O$_7$ the continued deviation of the VL form factor in the parent compound at the highest measured fields \cite{Cam22} could have a similar origin.  While the models constructed to describe the behaviour of the heavy fermion system, CeCoIn$_{5}$~\cite{Mic10, Dal11}, are not quantitatively appropriate for our results, we can draw several qualitative conclusions from them, and from more recent work \cite{Campillo2021}. Firstly, since we are at relatively small fractions of $H/H_{\rm c2}$, we expect the Pauli contribution to be small compared to the orbital component, leading to a deviation from the London model as we see here rather than the dominance of Pauli effects. Second, we expect the effect of Pauli paramagnetism to be larger in Ca-YBCO than the parent compound, since the effect is proportional to the effective mass which has been seen to be increasing with doping in this region of the cuprate phase diagram~\cite{Ram15, Put16}. This correlates with our observation here that the form factor in Ca-YBCO deviates from the London model at fields as low as 5~T, while in the parent compound it remained London-like until around 12~T, and further supports our earlier conclusion that calcium doping contributes holes to the system.

\begin{figure}
	\includegraphics[width=1\linewidth]{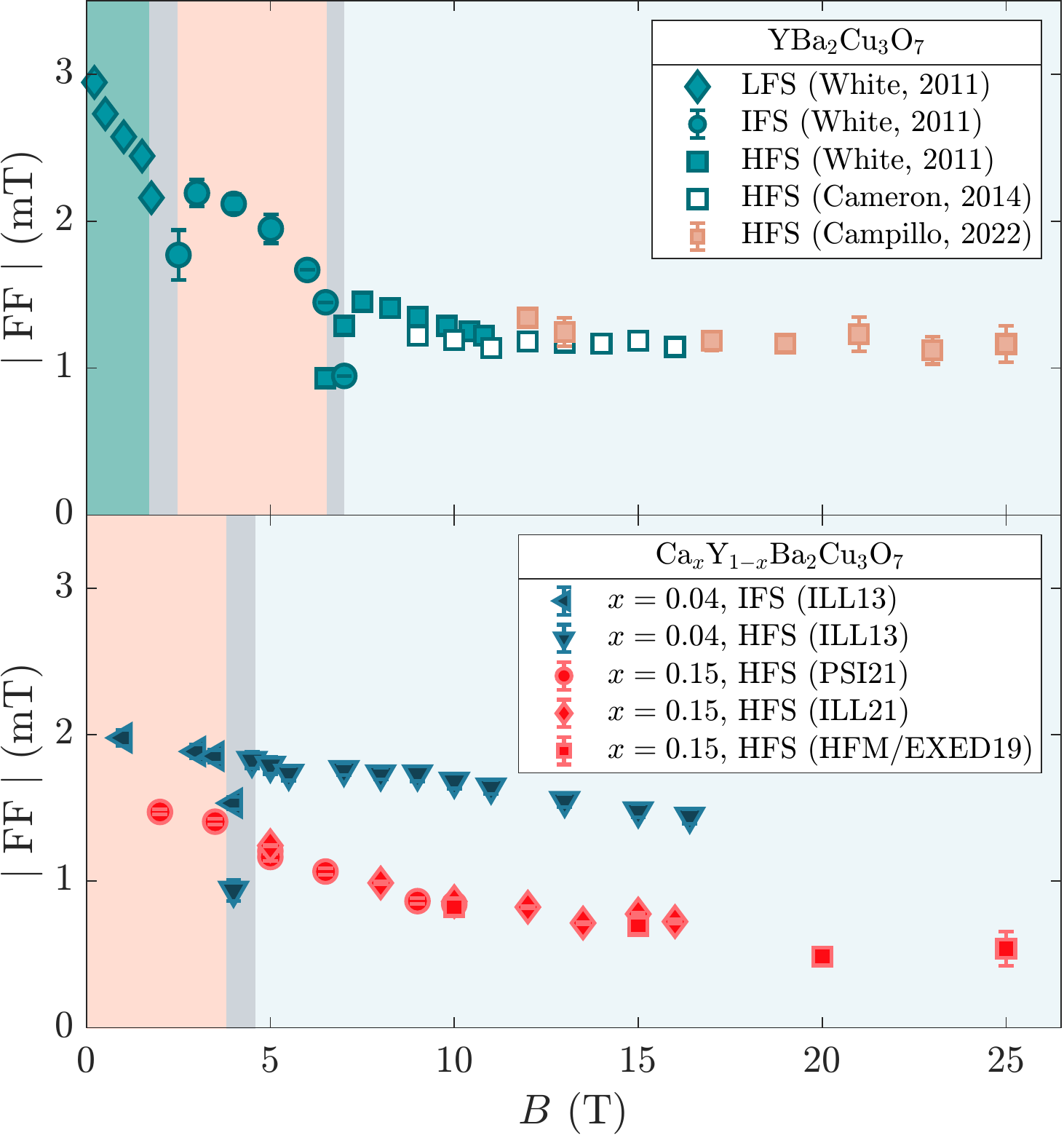}
	\caption{(Color online) Form factor of the VL as a function of applied magnetic field for YBa$_2$Cu$_3$O$_7$ and Ca$_{x}$Y$_{1 - x}$Ba$_{2}$Cu$_{3}$O$_{7}$.
	}
		\label{Graph5}
\end{figure}

\subsection*{Vortex lattice at higher temperatures}

\begin{figure}
	\includegraphics[width=1\linewidth]{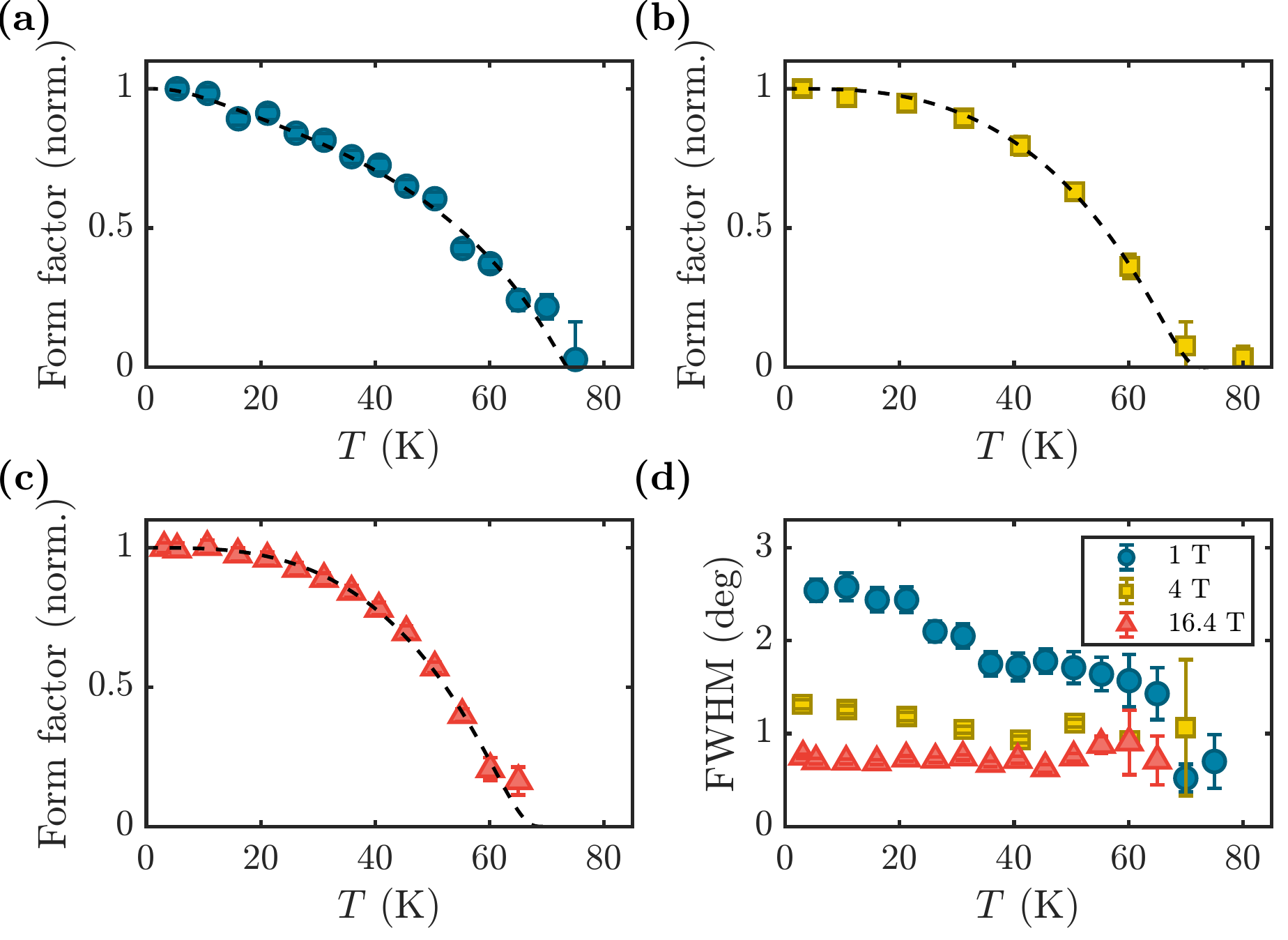}
	\caption{(Color online) Vortex lattice form factor as a function of temperature in an applied field of (a) 1~T, (b) 4~T and (c) 16.4~T. The solid line in panel (a) is a fit to the anisotropic London model using a \textit{d}-wave gap described in the text. Panel (d) shows the variation of the rocking curve FWHM as a function of temperature for the data presented in plots (a) to (c). }
		\label{Graph6}
\end{figure}

\noindent
We have measured the temperature-dependence of the VL signal in  4\% Ca-YBCO. The angle between the VL vectors was found to be temperature independent in these measurements, and the rocking curve width, shown in Fig.~\ref{Graph6}(d), remains reasonably constant with increasing temperature except at the lowest measured field of 1~T. This is in contrast to YBCO$_{7}$ at high field, which had a temperature dependence to the VL structure above an identifiable irreversibility temperature, which was visible both as a change in the angle between the VL vectors and the FWHM of the rocking curves~\cite{Cam14}. Passing above the irreversibility temperature was seen to reduce the static Debye-Waller factor, leading to a corresponding increase in the integrated intensity of the rocking curves. This rendered the temperature dependence of the VL form factor at high field unsuitable for fitting to models which had proved successful at lower fields and in other superconductors. The absence of both a strong temperature-dependent rocking curve width and of an identifiable irreversibility temperature suggests that the vortex lattice is strongly pinned by the effects of the Ca dopants.  We therefore expect that the variation of the VL form factor with temperature, which is shown in Fig.~\ref{Graph6}(a -- c), is not affected by changes in the perfection of the VL, allowing us to investigate the gap structure in this material.

We fit the temperature dependence of the normalised VL form factor to the relation in Eqn.~\ref{LondonFF}, and this is shown as the solid lines in Fig.~\ref{Graph6}~(a -- c). Following the method that has been applied in the modelling of the VL form factor in both cuprate and pnictide superconductors \cite{Whi11, Cam14, Fur11, Mor14}, we express the temperature dependence of the London penetration depth via an expression for the superfluid density:
\begin{equation}
\begin{split}
\rho_{s}(T) & = 1 - \\
& \frac{1}{4\pi k_{B}T} \int_0^{2\pi} \int_0^\infty \cosh^{-2} \left( \frac{\sqrt{\varepsilon^{2} + \Delta_{\textbf{k}}^{2}(T,\phi)}}{2k_{B}T} \right) \mathrm{d}\phi \mathrm{d}\varepsilon ,
\end{split}
\end{equation}
where $\rho_{s}$ is the superfluid density, normalised to its value at temperature  $T = 0$ and $1 / \lambda^{2} \: \propto	\: \rho_{s}$;  $\phi$ is the azimuthal angle about the Fermi surface and $\sqrt{\varepsilon^{2} + \Delta_{\textbf{k}}^{2}(T,\phi)}$ gives the excitation energy spectrum. The gap function was assumed to be separable into temperature and momentum dependent factors such that $\Delta_{\textbf{k}}(T,\phi) = g_{\textbf{k}}(\phi)\Delta_{0}(T)$, where $g_{\textbf{k}}(\phi)$ describes the momentum dependent gap as a function of angle around the Fermi surface. The temperature dependence of the gap, $\Delta_{0}(T)$, can be approximated by
%
%\begin{equation}
%\Delta_{0}(T) = \Delta_{0}(0) \tanh \left( \frac{\pi T_{\rm c}}{\Delta_0(0)}\sqrt{a \bigg ( \frac{T_{c}}{T} - 1 \bigg ) } \right),
%\end{equation}
%
\begin{equation}
\Delta_{0}(T) = \Delta_{0}(0) \tanh \left( \frac{\pi}{\alpha}\sqrt{a \bigg ( \frac{T_{c}}{T} - 1 \bigg ) } \right),
\end{equation}
where ${\Delta_{0}}$ is the magnitude of the gap at zero temperature and $\alpha$ \& $a$ are parameters related to the pairing state~\cite{Pro06}. The data were fitted to a \textit{d}-wave gap, $g_{k}(\phi) = \cos(2\phi)$ and BCS \textit{d}-wave pairing is represented by $\alpha =  2.14$ and $a = 4/3$. For larger values of the gap expected in high-$T_c$ materials, we retain the BCS temperature-dependence of the gap (values  of $\alpha$ \& $a$), but expect $\Delta_0 (0)$ to be of the order of $3~T_{\rm c}$~\cite{alpha}.

It has been previously noted in the cuprates that non-local effects can become increasingly important with field, which leads to a flattening of the temperature dependence of the superfluid density at low temperature. Work by Amin \textit{et al.}~\cite{Ami98, Ami00} showed that in the non-local regime, below a characteristic temperature $T^{\ast}$, the linear temperature dependence flattens out to a $T^{3}$ dependence. This behavior can be represented by the relation~\cite{Whi11}:
\begin{equation}
n_{s}(T) = 1-(1-n_{s}(T)) \left(\frac{T_{c} + T^{\ast}}{T_{c}} \right) \left(\frac{T^2}{T^2+(T^{\ast})^2} \right),
\end{equation}
where $n_{s}$ is the superfluid density, as calculated above in the local limit. $T^{\ast}$ is field dependent parameter given by $T^{\ast} \sim \Delta_0 ( \xi_0 / d) \propto \sqrt{H}$.

We employed the Ginzburg-Landau relation for the field dependence of the gap, $\Delta(B) / \Delta(0) = (1 - (B/B_{\rm c2})^{2})^{1/2}$, and a phenomenological relation for the critical temperature, $T_{\rm c}(B) = T_{\rm c}(0)(1 - B / B_{\rm c2})^{1/2}$~\cite{Tin96}. The fits used $B_{\rm c2} = 100$~T, which is a reasonable estimation for the upper critical field~\cite{Gri14}, and a critical temperature $T_{\rm c} = 76$~K, which is slightly lower than other reported values, but was consistent with all three sets of data we fitted.

We present the fits of the \textit{d}-wave model with non-local corrections to the temperature dependence of the form factor in Fig.~\ref{Graph6}~(a). Starting with the 1~T data in panel (a), we expect a $T^{\ast}$ of around 6~K from the BCS gap of $2.14~T_{\rm c}$, using the value of the critical field to estimate $\xi_0$. Leaving the magnitude of $\Delta_0$ as the free parameter of the fit, such that $T^{\ast}$ is determined by $\Delta_0$, we find that it returns a value of $\Delta_0 = (3.21 \pm 0.09)~T_{\rm c}$, which corresponds to a $T^{\ast} = 9.7$~K. While this gap value is $\sim 50 ~\%$ larger than the weak-coupling BCS value, we note that large gap values in cuprates are not unusual~\cite{Has14}, so this is a reasonable value. The fit is clearly a good description of the data, and possesses the finite low-temperature slope characteristic to nodal gap structures, indicating that this material is \textit{d}-wave in nature and that the non-local effects do not contribute strongly at low field, which is as expected.

For the fits at 4~T and 16.4~T, if we follow the Ginzburg-Landau relation for the field dependence of the gap, we would expect the gap value at 4~T to be $3.20~T_{\rm c}$ with $T^{\ast} \sim 20$~K, and at 16.4~T the gap to be $3.12~T_{\rm c}$ with $T^{\ast} \sim 40$~K. However, we find that these parameters are not able to fit the data. Indeed, following the predicted $\sqrt{H}$ dependence for $T^{\ast}$ does not allow for the fitting of the data for any value of the gap.
%, as the flattened low-temperature region characteristic of the contribution of non-local effects reaches too far up in temperature, and the fit always falls off too fast
 We have therefore kept the GL relation for the gap, which gives the values described above, but left $T^{\ast}$ as a variable parameter in these fits. This gives $T^{\ast} = 59 \pm 3$~K at 4~T and $T^{\ast} = 70 \pm 3$~K at 16.4~T. This is higher than expected, with $T^{\ast} \approx T_{\rm c}$ at the highest measured fields. This suggests that the onset of non-local effects is both more rapid and stronger with increasing field than the current models would predict.

\section*{Conclusions}

\noindent
We have investigated the vortex lattice structure of calcium doped YBCO and have compared the behavior of the vortex lattice with that of pure YBCO$_7$.  The results from the 15\% doped sample are dominated by pinning, but the 4\% doping shows similar behavior to YBCO$_7$ but with a lower field scale.  In the limit of high fields, both systems show the same vortex lattice structure.
%We observe that the field-independent high-field phase has the same structure in both systems.
%and that the temperature dependence of the VL form factor can in both systems be fit to a \textit{d}-wave gap at low fields with the inclusion of non-local effects at higher fields leads us to conclude that the underlying gap structure is the same in both systems.
We therefore conclude that the underlying gap structure is the same in both systems, and attribute the different field scale in the VL structural transitions mainly to the weakening of the 1-D superconductivity in the Cu-O chains by the disorder introduced by doping. Furthermore, in the 4\% Ca-doped sample the field dependence of the form factor can only be fit to the anisotropic London model with unphysically small coherence lengths.  This compares with the reported behavior in YBCO$_7$ where the anisotropic London model cannot be fit at high fields with any reasonable coherence length.  We speculate that this is due to the onset of a significant Pauli paramagnetic contribution inside the vortex cores as a function of field.

% we observe the contribution of a Pauli paramagnetic moment to the VL form factor, which appears at lower fields than in the parent compound due to an increase in effective mass of the carriers arising from the hole doping by calcium.

\section*{Acknowledgements}
This work is based on experiments performed at the Institut Laue-Langevin (ILL), Grenoble, and the Swiss spallation neutron source SINQ, Paul Scherrer Institute, Villigen, Switzerland and Helmholtz-Zentrum Berlin (HZB). EMF was supported by the Leverhulme Foundation. ASC acknowledges support from the German Research Foundation (DFG) under Grant No. IN 209/3-1. We would like to thank Robert Wahle, Sebastian Gerischer, Stephan Kempfer, Peter Heller, Klaus Kiefer and Peter Smeibidl for their support during the HFM/EXED experiment.

A. Cameron and E. Campillo contributed equally to this work.

%\bibliographystyle{APS_bib_v5.bst}
%\bibliography{Ca_YBCO_V2b.bib}


\end{document}